\RequirePackage{fix-cm}
\documentclass[runningheads]{llncs}            
\usepackage[numbers,sort&compress]{natbib}
\usepackage{graphicx}
\usepackage [english]{babel}
\usepackage [autostyle, english = american]{csquotes}
\MakeOuterQuote{"}
\usepackage{url}
\title{Why scholars are diagramming neural network models}
\titlerunning{Why scholars are diagramming neural network models}
\author{Guy Clarke Marshall\inst{1}, Caroline Jay\inst{1}, and Andr\'e Freitas\inst{1,2}}
\authorrunning{GC Marshall et al.}
\institute{Department of Computer Science, University of Manchester, Manchester, UK \\ \email{guy.marshall@postgrad.manchester.ac.uk, \\ \{caroline.jay, andre.freitas\}@manchester.ac.uk}
\and 
Idiap Research Institute\\
Rue Marconi 19, Martigny, 1920, Switzerland\\
}
\begin{document}

\maketitle

\begin{abstract}
Complex models, such as neural networks (NNs), are comprised of many interrelated components. In order to represent these models, eliciting and characterising the relations between components is essential. Perhaps because of this, diagrams, as "icons of relation", are a prevalent medium for signifying complex models. Diagrams used to communicate NN architectures are currently extremely varied. The diversity in diagrammatic choices provides an opportunity to gain insight into the aspects which are being prioritised for communication. In this philosophical exploration of NN diagrams, we integrate theories of conceptual models, communication theory, and semiotics. 

\end{abstract}




\section{Introduction}
\label{section:introduction}
Diagrams are a signifier, cognitive aid, and mediator of communication. In describing software systems, diagrams often provide a level of abstraction that facilitates an understanding of the overall structure, and the relation between the computational artifacts of the system. Software system diagrams have a dual role bridging between cognition and communication of humans, and representation of mechanisms entailed by machines. In the words of \citet{horn2002visual}: "Visual language has the potential for increasing human bandwidth, the capacity to take in, comprehend, and more efficiently synthesize large amounts of new information". 

In the field of AI, which feature many NN models, the pace of development is high, and as such conferences are the most prestigious academic venues. In their scholarly proceedings, we find that the majority of papers include a system architecture diagram by way of structural explanation. This is common across Computer Science and other model-centric domains. What is less common is the variety of these representations, even when compared with other research and engineering domains. Despite being based on similar and mathematically-well-formalised computational artifacts, such as neural networks, the diagrams have very low consistency. In this paper, we utilise the opportunity provided by the lack of convention to gain insight into the way the creators of NN models are choosing to communicate their models. This follows the recent thinking of \citet{even2021lines}, who explores medieval scholastic diagrams and the insight they provide into the thinking of their creators.



This study is motivated by a number of questions. Why are diagrams being used to describe NNs? Why are the diagrams so heterogeneous? Progress is made to this end through drawing together recent empirical experiments \citep{marshall2020understanding,marshall2021corpuslong,marshall2021structuralist,marshall2020researchers}, supported by the following synthesis and reasoning: (i) There is a relation between content included in NN diagrams and their role in the scholarly community as conceptual models, and (ii) There are visual encoding prioritisations which align with subcategories within mental models theories.

\section{Background}

AI is software, written in a programming language, and often based on neural networks. A neural network takes an input (usually text, numbers, images or video), and then processes this through a series of \emph{layers}, to create an output (usually classification or prediction). Each layer contains a set of \emph{nodes} which hold information and transmit signals to nodes in other layers. Specific mathematical functions or operations are also used in these models, such as sigmoid, concatenate, softmax, max pooling, and loss. Different architectures are used for different types of activities: Convolutional Neural Networks (CNN), inspired by the human visual system, are commonly used for processing images, and Long Short Term Memory networks (LSTM), a type of Recurrent Neural Network (RNN) which are designed for processing sequences, are often used for text. These neural networks "learn" a function, but have to be trained to do so. Training consists of providing inputs and expected outputs, so the model can learn how a function should be inductively represented. The model is then tested with unseen inputs, to see if it is able to process these correctly. 
The explicit data perspective (focused on vectors and matrices) and the functional perspective are fairly distinct ways of thinking about what is happening, which is part of the problem of communicating in this area. The architectural perspective on the model can be to a greater or lesser extent encompassing the data manipulations, which leads to a broad spectrum of possible representations. 

Neural networks can also be considered as a set of transitions between latent states (a different way of conceptualising the data). These latent states are fundamental to the nature of NNs, which often utilise mechanisms such as gradient descent via back-propagation which are sufficiently conventional to be omitted from non-pedagogical diagrams. The functions are not highly heterogeneous, but the representations are heterogeneous. 

The diagrams are a sequence of latent states with attributes associated with the latent space. The models themselves are comprised of different types of layers and different sizes or parameters for the layers, which are usually encoded in diagrams. The configuration, dependencies between latent dimensions, and attributes of layers are included as the diagrams are utilised by authors attempting to produce a description of these latent states using graphical components. 

The mathematics of NN's are well established and have an established low level representational language. At the system level, diagrams are used but conventions are not well established.

There are fundamental and complex aspects of NNs which are not commonly included in diagrams. Authors put substantial thought into the nature of the loss function, and how to decompose the set of components of the loss, but this is done outside the diagram and articulated through mathematical notation. It is integral to the NN, but the design choices for layers are prioritised in the diagram.

In terms of communicating about NNs, it is not just \emph{from researchers to the public}, but also \emph{between researchers} that attention is needed. The act of communicating between researchers is not only about making code reproducible \citep{Hutson725}, but also making cognitively accessible the models and the computational artifacts of which they are comprised. As a result of this, diagrams have a role to play in the ethics of NNs and AI more broadly, and are a key way of establishing transparency and managing risks around how the model operates.

\section{Diagram content relates to conceptual models}
We claim there is a relation between content included in NN diagrams and their role in the scholarly community as artefacts of conceptual models. To explore this, we first describe the heterogeneity in representation, and then proceed to classify the observed phenomena, linking the observed diagrams to conceptual and mental models. 
\subsection{Heterogeneity in representation}
\label{section:heterogeneity}

In principle, the primary purpose of diagrams in scholarly publications is communicative. The authors are attempting to communicate through a diagrammatic medium some kind of relational structure. Diagrams are ideally suited to this task, and are used for this purpose in many domains \citep{ma2020domain}. However, there are other social aspects. Without passing too critical an eye over the scientific endeavour, having a "good looking" diagram (both aesthetically and technically) may improve perception of a paper, thereby making it more likely to pass through peer review. While the visual encoding methods are quite unconstrained and heterogeneous, it is conventional to include an architecture diagram if the paper presents a novel model. The subsequent discussion attempts to make steps towards understanding the reasons for, and consequences of, this heterogeneity.




Perhaps due to the complexity of NNs, there are a number of diagrammatic representational choices that are made by different authors attempting to express different things. This has been shown by \citet{marshall2020understanding}, who use VisDNA, a grammar of graphics \citep{engelhardt2020dna}, to demonstrate the heterogeneity of visual encoding principles employed in this domain.

In an interview study, \citet{marshall2020researchers} identify heterogeneous use cases and preferences associated with NN diagrams. The modal use case mentioned by all participants was ``how the system works''. Confusions reported were around the ``flow'' of reading the diagram, the purpose of the system, and gaps or lacks of specificity within the diagram. The interview study also found a huge range of diagramming tools are used to create these diagrams, and identified themes in usage. Three major themes were identified, covering visual ease of use, appropriate content and expectation matching. ``Visual ease of use'' related to clear navigation, aesthetics, consistency within the diagram and having distinct process stages. ``Appropriate content'' referred to either wanting more or less information in a diagram, or preferring multiple different diagrams to display different information. These contradictions within these themes reflects the different priorities users had for specificity or an instantiated example. ``Expectation matching'' found that users want consistency across the diagram and within the domain, and to have symbols or abbreviations explained.

Given the social nature of research, it is curious that there is not more prevalent "copying" or adoption of informal diagrammatic encoding conventions, or even convenience-based similarity caused by the use of popular diagramming tools. Given that it would be practically easier for authors to directly copy existing styles, it is unlikely to be chance, but rather we argue that there must be a compelling reason for authors to be creating such different diagrammatic representations. 

A partial explanation for heterogeneity could be a lack of appropriate diagramming tools. In a recent interview study involving technical domain experts, \citet{ma2020domain} found that "To illustrate concepts effectively, experts find appropriate visual representations and translate concepts into concrete shapes. This translation step is not supported explicitly by current diagramming tools". This does not explain why NN diagrams are so heterogeneous compared to diagrams in other scholarly domains, nor does it explain the lack of informal conventions. 

\citet{nefdt2020puzzle} argues that the state of understanding of AI systems is such that "meaningful components" and compositional structures have not been established for deep neural networks. He does not extend his argument to the entire system, focusing instead on the mathematical "black box" that is a deep neural network. If we adopt this stance over an increased scope, this would seem to support the claim that there is an underlying cognitive reason for the differences in representations. As such, we argue that insight into the current state of understanding can be gained partially \textit{as a result of} the current "epistemic opacity", and therefore allows an insight into the representational priorities of the author.

We hypothesise that a particular representational aspect is prioritised by the author either because it shows what they think is important, or because it is what they would want to see in a diagram authored by their peers. In either case the priority is effective communication. Differences in prioritisation may be causing the creation of bespoke diagrammatic encodings. When representing NNs, the diagram author can prioritise different aspects of what is represented:
\begin{itemize}
    \item Function: Operations which occur, representation transformations, and the purpose of parts of the model
    \item Data: The data model, type, dimensionality and how it is manipulated
    \item Example: Showing the steps of an example input through that model
    \item Contribution: Focusing on the scientific novelty of the approach, giving much more detail in that area
    \item Code: Important class names and the order in which they are called
    \item Mathematics: Including specific mathematical functions
    \item Index to text: Using a label structure to allow for easier referencing
\end{itemize}
These representational priorities result in different content being displayed through different visual encoding mechanisms. It is usual to have aspects of several of these priorities, as it is not the case that the prioritisation of one aspect necessarily inhibits another. 
In terms of how the diagrams are presented within a paper, some papers include multiple diagrams, either by multiple figures or sub-figures. Often, sub-figures or boxes are used to give both schematic and detailed views within the same diagram. Dependencies are often indicated by arrows. Diagrams almost always represent important content in natural language, such as labels or descriptions.



\subsection{The content of diagrams}


The inclusion of an "example" makes the diagram be "of" a particular run of that model. However, it is understood that this is signifying the model itself. There are some diagram users who take an example instantiation and use this to generalise to the operation of the overall model \cite{marshall2020researchers}. Indeed, in this way the diagram is supporting logical induction, rather than deduction from general rules (which would be more classically descriptive of Function). Similarly the omission of inputs and outputs (e.g. "text" or "probability") from the diagram makes the diagram be "of" a set of operators rather than of a functioning model, which is equally understood as signifying the model (indeed, including data).

Rarely are NNs' Knowledge Representations (links with semantics) represented visually in this scholarly community. Instead, labels are favoured for this important component. In some diagrams, they include the dimensions of the embedding space, a drawing of an arbitrary graph, or a simplified example of part of the knowledge representation. The high dimensionality and the sparsity involved makes visual representation challenging.

\subsection{Specific graphical objects}
Arrows have been linked with the concept of functional processes \citep{heiser2006arrows}. All surveyed diagrams used arrows, perhaps suggesting that the models are understood as being compositional and sequential (rather than as objects which enact a function).

Labels for "layers", sets of neural nodes often performing a particular neural network function, are common. These NN functions often have metaphorical or descriptive names, such as "embedding" or "attention", which form an informal lexicon of computational artefacts. The labels are often used alongside more complex sets of graphical objects signifying the NN, performing the role of (linguistically) simplifying and linearising the model without taking a sentential form. To take a specific example, the layer labels are often of the form "embedding layer", i.e. mentioning the function. This in turn might be labelling a set of blocks labelled "BERT". In this sense, this type of diagram contains cognitive support for two parallel mental models, that of function in the original diagram, and that of state in the layer labels. This particular aggregation-and-switching-mental-model usage of labels is very common, but not exclusive. Some diagrams may instead have mathematical functions (e.g. "tanh") labelling layers. Principally the association between these representations is grouping by alignment, but may additionally make use of "linking" graphical objects such as arrows, brackets, or blocks, or grouping by colour. Aspects of the content were surveyed by \citet{marshall2021structuralist}, who propose groupings for NN diagrams based on their ``diagrammatic'' (relational) and ``content'' (semantic) features. The four proposed NN diagram classes are Visual, Mathematical, Lightweight and Unorthodox. Challenges associated with creating groupings of these diagrams are noted, and further comment: ``We have not shown that this heterogeneity is a problem per se, but it does differ from many other Computer Science domains where diagrammatic representation of systems is more standardised.''

Many graphical objects, and almost all diagrams, utilise labels (in English). This suggests a low level of semantic content within the graphical objects. Perhaps this is because of the non-physical nature of the models, which makes it challenging to signify visually without the aid of an established formal language.

\section{Visual encoding may relate to conceptual modelling}
\label{section:mentalmodel}
This section aims to link diagramming priorities with conceptual models, as has been speculated in historical domains \citep{even2021lines}. This link could serve to explain some of the manifest heterogeneity.

\subsection{Related work}
\citet{Clark1998TheMind} proposed the ``Extended Mind Theory'' which encompasses diagrams, proposing that diagramming shapes (and is shaped by) cognitive processes. This has been followed by a multitude of successive theories. \citet{even2021lines} takes this forward in her recent book ``Line of thought: Branching Diagrams and the Medieval Mind'', applying it to medieval scholastic diagrams sometimes found in the margins of books. These tree diagrams are unformalised, sometimes being readable as multiple sentences, and sometimes forming concept trees or hierarchies. Both empirical corpus-based analysis and theoretical reasoning are employed to understand phenomena. The diagrams' role is postulated to be about summarising and making sense of the work. Even-Ezra is it pains to argue that these medieval diagrams are not solely mnemonic devices, though they are sometimes viewed by contemporary scholars as such. She concludes (p49) that ``the best conclusion regarding the technique's functions and purposes, however, is that there is no one-size-fits-all answer.'' We reach a similar conclusion regarding present-day NN diagrams, where the externalisation facilitated by diagramming should attempt to fit the author's reasoning needs and communicative aspirations. It is also notable that diagrams are neglected from scholarly discourse about NNs. Some scholars of science education (e.g. \citet{ainsworth2011drawing}) advocate the use of diagrams in learning, rather than solely as mnemonic devices.

Many scholars examine the effectiveness of diagrams and the relation of this to cognition. The Cognitive Dimensions framework \cite{blackwell2003notational} is designed to support user interface design, including paper-based notational systems, and to provide a broad-brush evaluation tool. It does not attempt to be an analytic method, but a discussion tool for designers. As such, it operates at a fairly high level. The framework includes task-specific attributes such ``viscocity'' (the number of actions required to achieve a goal) and ``Hard mental operations: high demand on cognitive resources''. By design, the attributes are not necessarily empirically tractable: Cognitive Dimensions are often challenging to measure even for a well-defined task, facilitating discussion about trade-offs between e.g. reducing viscosity by increasing abstraction level, at the expense of increased mental operation. This is by design - the framework provides broad-ranging prompts to allow critical consideration of designers' notational systems. The framework has been widely applied in many areas of software, including in HCI \cite{dagit2006using}, as part of the study of API usability \cite{piccioni2013empirical}, and even extended to examine the role of conceptual visualisations in collaborative knowledge work \cite{bresciani2008collaborative}. 

Cheng \cite{cheng2016constitutes} provides a taxonomy of 19 cognitive criteria for analysing representations. It would be possible to analyse NN diagrams with Cheng's framework. ``A1.1 One token for each type", for example, is violated within many diagrams and within the corpus, where tensors are represented in a plethora of visual encodings. However in the present domain much groundwork is missing. For example, to discuss suitability of diagram components against reading and inference operations such as ``A2.2. Prefer Low Cost Operators" appears to require domain-specific evidence which has not yet been established.

\subsection{Mental models and conceptual models}
We can further refine this schema to correspond more directly to a theory of mental models. Definitions of mental models are various, but can perhaps be summarised by: "A mental model is a simplified representation of reality that allows people to interact with the world." \citep{jones2011mental} 

In Computer Science, \citet{guarino2020philosophical} recently developed a model for the relationship between mental models and conceptual models. For \citeauthor{guarino2020philosophical}, mental models are "personal, partial accounts of the external reality, filtered through the lens of a conceptualization, that people use to interact with the world around them." A mental model (or perhaps less ambiguously a conceptual mental representation), is described by a conceptual model, in our case the diagram with a communicative purpose. "Note that, being an information object, a conceptual model is always the result of an intentional act. In other words, conceptual models are artifacts produced with the deliberate intention of describing a conceptualized reality." With further specific reference to the Computer Science domain, \citeauthor{guarino2020philosophical} state that "we may say that a computer program is a conceptual model of the computer’s internal behavior, but only as long as its programming language’s primitives denote concepts concerning computer behavior. If they rather denote data, we conclude that such a computer program is not a conceptual model." \citeauthor{guarino2020philosophical}'s work highlights the challenges of philosophical precision in computer science when discussing mental models. Applying these definitions to NN diagrams, it is variable whether the diagram is a conceptual model. As such, in this work we use mental models as a metaphor, rather than in the strict sense. Usage of the term in this manner is common practice in human-computer interaction. 


To summarise our working definitions: The \textit{conceptual model} is the artefact itself, with communicative purpose, to articulate concepts and the relationships between concepts. Diagrams are often used for this purpose, as an artefact representing the creator's conceptual model, and have been found to aid learning and reasoning \citep{ma2020domain}. Note that this definition fits with Extended Mind Theory, where the artefact is viewed as embodied cognition \citep{rowlands2010new}. In our usage, a \textit{mental model} is the way an author is thinking of the model when creating that conceptual diagram for use in scholarly publication. We draw on psychological mental model research, itself utilising a more rigorous definition, in order to assist in the classification of conceptual diagrams, because the fit appears natural. We explore potential consequences if this is aligned. However, we do not suggest a deep claim about mental operation or cognition, which would require an empirical study. 





\subsection{Classification using mental models theory}

\begin{figure}
    \centering
    \includegraphics[scale=0.3]{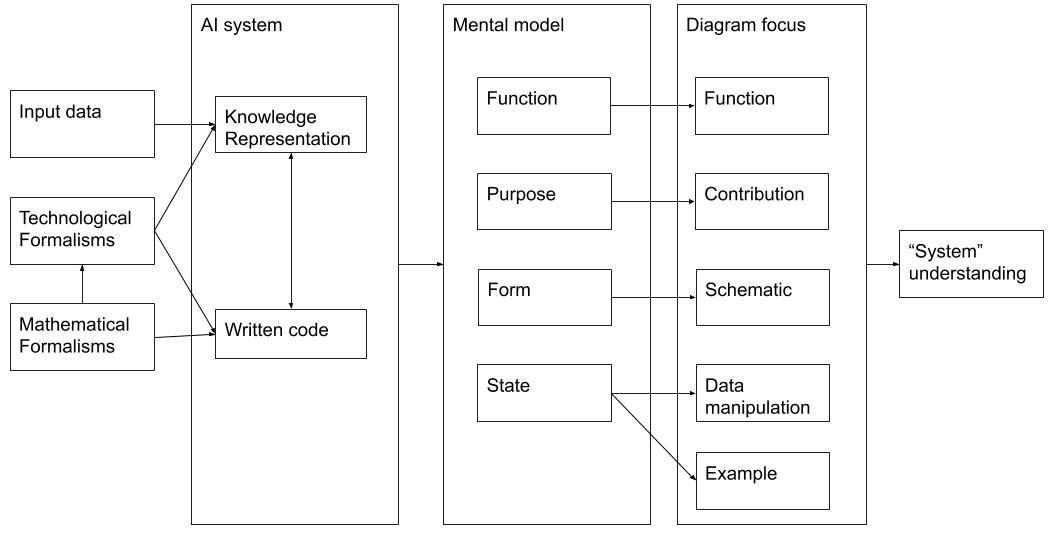}
    \caption{Representational choices in diagrams of NNs}
    \label{fig:representations}
\end{figure}

In this work, we are focusing on the diagrammatic metaphor being used for \emph{what is represented} rather than the visual encoding, in order to provide a different perspective to analysing NNs through the lens of VisDNA \citep{marshall2020understanding}. Bringing types of NN diagrams closer to mental models, we can draw parallels with mental model categories \citep{rouse1986looking} and types of diagram observed. Diagrams may have aspects of multiple categories.

\begin{itemize}
    \item Function: Explaining how the model operates by emphasising functional aspects, such as mapping, input and output. Operations used as a verb. For example, "word embedding" (a general term) rather than "BERT" (a specific architecture for embedding). This type often includes example input and output. 
    \item Contribution (Purpose): Omits the majority of information other than that required to understand the sub-section of the model that contains the novelty of the model or approach. 
    \item Schematic (Form): Describes the model at a high level, uses probably a block-style without iconic graphical objects. Does not include mathematical or data details. The schematic may relate to classes or packages used. In order to be distinct from Function, it commonly uses e.g. "BERT" rather than "word embedding". Also often aggregates graphical objects into modules. (In order to be "state" (i.e. what it is doing) the block diagram should be verbs).
    \item Data manipulation (State): Includes data dimensions, and usually a visual representation of the data itself. It describes what the model does to the data, so this also includes where operations are primarily labelled arrows (rather than inside blocks).
    \item Example: How example data transforms. Includes example input, output and intermediate steps. Usually better relates to Function (how it operates) rather than Form (what it looks like), but this depends on the graphical objects used. It is useful to disambiguate this, particularly for Image Processing, where often the diagram is a visual representation of the data manipulation. In one sense this is the Form of the example data manipulation, and in another sense the Function of the model on a single example. It is the latter that we are concerned with, in our assertion that the diagrammatic representation is signifying the model rather than the example. Note also that the inclusion of intermediate step as using the same example is important. Many diagrams include example inputs and outputs, but in the processing of the model they do not utilise the example and the diagram can be "schematic"). 
\end{itemize}



\begin{figure}[htbp]
    \centering
    \includegraphics[scale=0.25]{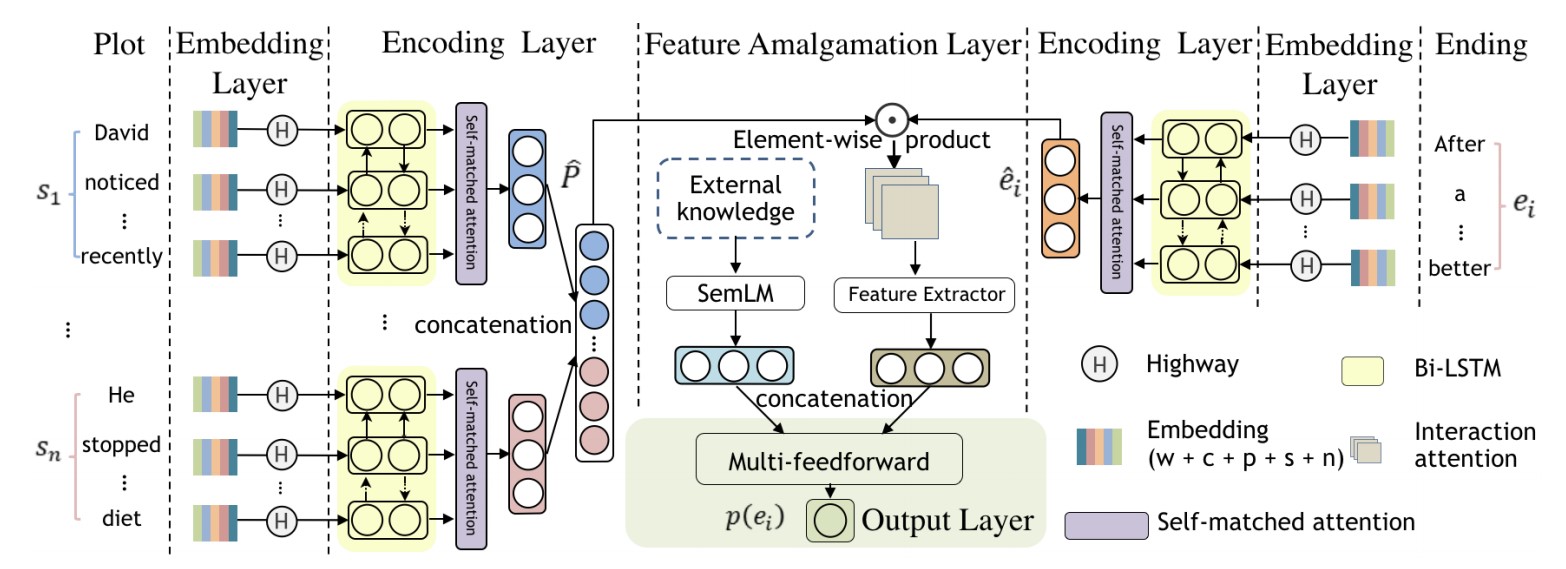}
    \caption{An example diagram, used by \citet{li2018multi}, using labels for Form, graphical objects variously for Form and State, input and output for Purpose}
    \label{fig:example}
\end{figure}

Sometimes the diagrammatic representations found in conference proceedings contain errors. In addition to typographic errors, these can be visualisation errors, in the sense that the diagrams may cause confusion or inaccurately reflect the reality of the underlying model. For example, the circles representing vectors can represent a precise number of objects, or not. Fig. \ref{fig:example} shows an example where the pairs of circles represent two LSTM output vectors (a common representational choice), while the three circles of $\hat{P}$ do not represent three vectors but rather $j$ vectors, where $j$ is the number of words in the sentence. The omission of the ellipsis following the embedding layer appears to have led to this visual encoding choice. This unmeaningful 3-vector is repeated perhaps more dangerously in the final concatenation before "multi-feedforward". This can be understood by careful reading of the words and formulae in the text, but could be misleading, as found in an interview study by \citet{marshall2020researchers}.

We claim these potential different forms of diagrammatic representamen are signifying the same referent, a NN model, through different lenses. We are not claiming that precisely the same sense is made with different diagrams. Even strictly isomorphic representations have different semiotic properties, such as the perceptual and cognitive properties of mathematics as performed by equation, Euler diagram or language, or indeed without the ability to perform gestures.

The separation of Form from Function is neither self-evident nor entirely natural. In some software paradigms, such as functional programming, the essence of Form and Function are entangled. By being faithful to the perspective of the model's Form and Function, rather than to the human cognitive process which it is automating, we are able to be crisper with this distinction. 

The types of representamen employed are of particular interest. Our hypothesis is that the diagrams, and the variety we see exhibiting the above principalities, are a result of the range of cognitive functions being employed by different users. At present, there is no common language they are using to communicate. No suitable representation providing as cognitive support such as symbolic equations gave to the mathematicians of ancient Babylon, or the letter $x$ gave to \citet{GlossarWiki:Descartes:1637}, has been sought nor found. In Physics, \citet{coecke2017picturing} use diagrams to reason about quantum processes and to diagrammatically perform the calculations required to understand them. 

Another important aspect is reproducibility. Circuit diagrams and other standard diagrammatic representations, often implemented or overseen by professional bodies, have also enabled this standard form of communication and reproducibility across many domains, including Computer Science. With engineering diagramming technology in mind, we proceed to consider specific graphical objects used in NN diagrams.

\subsection{Mental operations}
\label{section:mentaloperations}
Ryan Tweney's lifetime of work examining Faraday's cognitive processes through images provides insight into a single author's cognitive process (see \citet{ippolito1995inception}). Without the breadth and depth of work of a single author, it is not practical to carry out a similar analysis in NNs. Nevertheless this work served to inspire the approach taken here, and necessitates brief discussion of "mental models" in the cognitive science sense.

All three primary mental operations of apprehension, judgement, and inference \citep{hobhouse2013theory} are at play in using NN diagrams for research. 
Apprehension, being the forming of a picture in one's mind, is important for depicting and understanding of the model. Any example of Judgement could be "this is relevant", and Inference "this is a contribution". We focus on the creation of mental representation, i.e. Apprehension. Note that "Mental models can be constructed from perception, imagination, or the comprehension of discourse" \citep{johnson1983mental}, and conference proceeding NN diagrams are in some sense all three of these: Perception of the diagram, imagination of the model operation, and comprehending scientific discourse. 

We can categorise mental models as being about Purpose, Function, State and Form \citep{rouse1992role}. These categories of mental models can be related to the types of diagrams found in NN literature, as shown in Fig. \ref{fig:representations}. As highlighted by \citet{rouse1986looking}, these facilitate describing, explaining and predicting: “Mental models are the mechanisms whereby humans are able to generate descriptions of system purpose and form, explanations of system functioning and observed system states, and predictions (or expectations) of future system states.” Following this, we have the following categories applied to NNs:

\begin{itemize}
    \item Purpose is why a model exists, specifically the activity it performs. There are canonical tasks often linked to human cognitive operations (such as Named Entity Recognition or Image Classification). The discipline is evolvable and models are comparable because there are these canonical tasks. \textit{Manifestation}: Often evidenced by inputs and outputs. Title, caption, other language or images describing the task. 
    \item Function is how the model operates. \textit{Manifestation}: The inclusion of model operations as category terms. Graphical objects representing nodes. Mathematical equations. 
    \item State is what the model is doing, with temporal elements. In NNs, this is about data states at a particular functional position. \textit{Manifestation}: Often shows how the dimensions of the data are changed, and usually a visual representation of the data itself. It describes what the model does to the data, so this also includes where operations are primarily labelled arrows (rather than inside blocks).
    \item Form is what the model looks like, and how the model is arranged. It is succinct, expressive and general, comprised of the representational and functional choices, how they are structured and composed, with emphasis on the dependencies between components. The physical structure analogy is perhaps closest to the classes and modules of code, and the shape of the data. This is usually succinct. \textit{Manifestation}: The inclusion of components as nouns. Often aggregating into labelled modules. 
\end{itemize}

A ``State Diagram'' is sometimes used in Computer Science to describe a system with a finite number of states and their transitions, is available in a standardised form in UML \citep{Booch1998UnifiedThe}. Whilst theoretically this could be used to describe many NNs, empirically this is a very uncommon scholarly choice of representation. 

At first glance, it may seem that the correspondence between mental models is not strong for the "Function" model. As noted previously, in what is observed some diagrams prioritise an instantiated example. This can be interpreted as a different way of reasoning (inductive rather than deductive), and in both cases appears to be focused around articulating Function. The detail-centric vs function-centric comparison partially describes how abstract or schematic the diagram is. A difference might be the inclusion of a block labelled as "CNN", only including hyperparameters in the former. The inclusion of hyperparameters, which are non-structural configuration details, demonstrates that the diagram supports a different level of reasoning.

\subsection{Abstraction levels}

\begin{figure}
    \centering
    \includegraphics[scale=0.35]{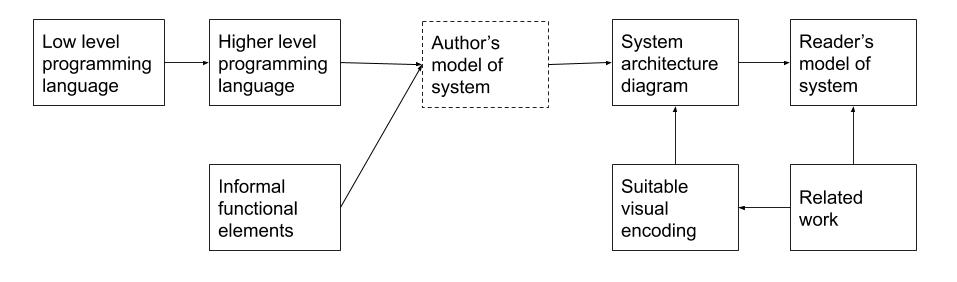}
    \caption{Abstraction levels used in interacting with NNs}
    \label{fig:abstraction}
\end{figure}
Diagrams can be used to represent information at any level of abstraction. Fig.~\ref{fig:abstraction} shows part of the context of abstractions and formalisms surrounding architecture diagrams, particularly in the absence of a formalism to describe functional modules, or specified visual encoding frameworks. It is quite possible that the model creator does not have a detailed knowledge of how their code will execute (or even where, in the case of cloud computing). Of course, mathematical formalisms and related work may also be expected to form part of the author's model of the NN. With the complexity of NNs, it is unlikely to be practical to perceive both the granular underlying mathematics and the overall NN structure in one instant. Indeed, the structural explanation given by a diagram is necessarily schematic, omitting information in order to be more efficient. In Fig.~\ref{fig:abstraction}, each of the arrows has some loss. Not all aspects of the author's model are included in the architecture diagram, for example. 

In a single diagram, there can be different levels of granularity. Graphical representations of vectors are employed alongside higher level routines such as biGRU or LSTM, themselves often comprised of multiple vector operations. The different abstractions and granularity levels available to diagram authors facilitates the diversity of representations that are observed, both within a single diagram and across the diagrammatic corpus.


\subsection{Diagram formats representing the same model}

\begin{figure}
    \centering
    \includegraphics[width=0.7\textwidth]{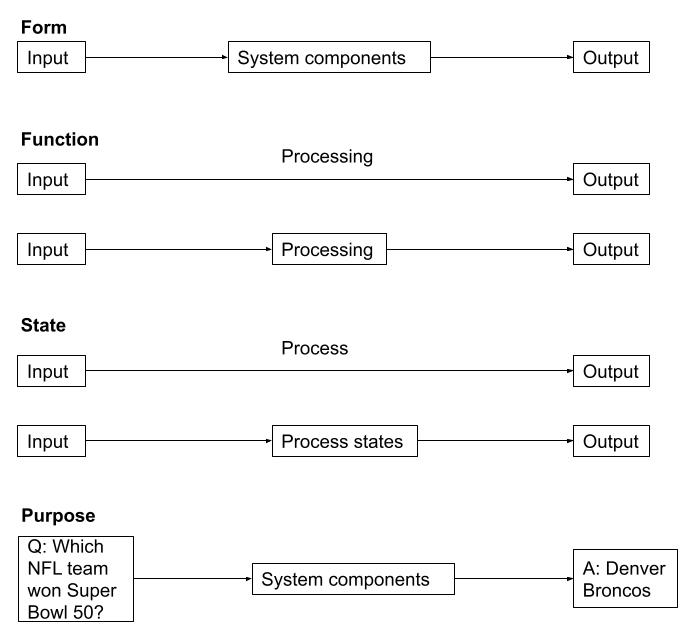}
    \caption{Example representations of the same model, obfuscating detail and graphical encoding differences}
    \label{fig:schematics}
\end{figure}
Fig. \ref{fig:schematics} shows examples of the overall structure of diagrams commonly found in NNs. These schematics omit many details and complications of "real" diagrams. See Fig. \ref{fig:example} for a typical example from ACL 2018, noting the lack of linearity and variety of visual encodings used even within a single diagram. Whilst the two versions of Function and State represent equivalent content (in that there exists an isomorphism to convert between them), they are not identical. Each structure places emphasis on different aspects of the representation. 

We can apply this framework to real examples. Fig. \ref{fig:example} shows a diagram taken from proceedings of a top conference, describing the architecture that is their contribution. The diagram has labelled layers describing Form, limited Purpose (example input and output, with no intermediate steps, but the labels of "plot" and "ending" facilitate abduction), and the majority of diagrammatic real estate given over to graphical elements describing vectors (the circles).  Independently of the text, abbreviations such as $s_i$ may be understood by their context and abduction. Concatenation is a (nominalised) verb on arrows, while Feature Extractor, element-wise product, and the majority of other labels are nouns. Combined with the emphasis on the data/vectors themselves, although they do not have dimensions indicated either numerically or pictorially, it would seem this diagram is primarily State-based, with Form labels. We hypothesise the author is thinking about the model as manipulating data, as they prioritise communicating State but without much information about the data itself (just grouping by colour, and repetition for scale). This suggests that the author feels the "important" part of the model to communicate is not necessarily the way the data changes but the overall model. We hypothesise this "perceived importance" is in itself a window into the mental representation of the authors: In the absence of any independent advice into effective diagrams, authors using "examples" to instantiate their cognition about a model will see it as important for others to include "examples" in their own diagrams.

\section{Conclusion}

We argue that the heterogeneity in diagrammatic representations of NNs is due to the inherent complexity of what is being represented and a lack of obvious good representational choice for the model themselves. In the case of many other scientific and engineering disciplines, a standard has quickly emerged. It may be that we are too early in the life of "NN science and engineering" to see this, and instead are able to gain insight from the heterogeneity. The heterogeneity seen today is a manifestation of a lack of conventional model elements and visual encoding principles. 

NNs are a new medium, which at its most granular representational level are not easily interpretable. Currently there is heterogeneity in diagrammatic representation. Diagrams are a useful, and efficient, way of understanding NNs. Diagrams are a fundamental signification layer, encoding the design and meta-description of these emerging models. We have linked this to a morphic semiotic perspective: "code interpretation" being more computationally expensive than "diagram interpretation". We have hypothesised the diagrams used at present to have relationship to mental models. As the community evolves, and the representational requirements of these diagrams become clearer, as with many languages, we might expect some consolidation around effective ways of signifying the underlying model, and providing cognitive support for communicating and reasoning about NN models.  




\renewcommand{\bibsection}{\section*{References}}
\bibliographystyle{splncs04nat}      
\bibliography{bib}   

\end{document}